\documentclass[conference]{IEEEtran}
\IEEEoverridecommandlockouts

\usepackage{cite}
\usepackage{amsmath,amssymb,amsfonts}
\usepackage{algorithmic}
\usepackage{graphicx}
\usepackage{stfloats}
\usepackage{float}

\usepackage{booktabs}
\usepackage{textcomp}
\usepackage{xcolor}
\usepackage[hidelinks]{hyperref}

\def\BibTeX{{\rm B\kern-.05em{\sc i\kern-.025em b}\kern-.08em T\kern-.1667em\lower.7ex\hbox{E}\kern-.125emX}}

\begin{document}

\title{Multi-Agent Orchestration of 3GPP Channel Estimators}

\author{\IEEEauthorblockN{I. Zakir Ahmed and Hamid Sadjadpour}
\IEEEauthorblockA{Department of Electrical and Computer Engineering\\
University of California, Santa Cruz}}

\maketitle

\begin{abstract}
Pilot-aided channel estimation is a decisive block in orthogonal
frequency-division multiplexing (OFDM) receivers for both 5G New Radio (5G-NR)
and Long-Term Evolution (LTE). A large body of estimators exists---from simple
least-squares (LS) interpolation to statistically optimal linear
minimum-mean-square-error (LMMSE) variants and, more recently, deep
convolutional denoisers---yet no single estimator is uniformly best: the winner
depends on the propagation scenario, the numerology, the operating
signal-to-noise ratio (SNR), the mobility (Doppler), and the antenna
configuration. In this paper we quantify this fact through a unified study of
eight literature estimators evaluated over the 3GPP TR~38.901 Urban-Macro
(UMa), Urban-Micro (UMi) and Rural-Macro (RMa) channels generated with NVIDIA
Sionna, for both 5G-NR and LTE numerologies, in single-input single-output
(SISO) and $8\times2$ multiple-input multiple-output (MIMO) settings. We then
propose a \emph{condition-adaptive multi-agent orchestrator} that treats each
estimator as an independent agent and dispatches, per operating condition, to
the agent that is best on a validation split---without any genie knowledge. The
orchestrator tracks the per-realization oracle to within $1.07$~dB and improves
the normalized mean-square error (NMSE) over the best \emph{fixed} strategy by
up to $3.6$~dB at high SNR, where the low-SNR champion is no longer optimal.
Because the agents are independent, running them concurrently delivers this
best-of-eight accuracy at essentially single-estimator latency: a data-parallel
partition scales the wall-clock nearly as $1/K$ with $K$ workers (up to
$6.9\times$), whereas naive by-algorithm partitioning is Amdahl-limited by the
heaviest agent. The results substantiate multi-agent orchestration as a
practical route to robust channel estimation across heterogeneous 5G-NR/LTE
deployments.
\end{abstract}

\begin{IEEEkeywords}
Channel estimation, OFDM, 5G-NR, LTE, MIMO, LMMSE, deep learning, 3GPP
TR~38.901, multi-agent orchestration.
\end{IEEEkeywords}

\section{Introduction}
Coherent detection in OFDM systems requires an estimate of the channel
frequency response (CFR) at every data subcarrier. In 5G-NR and LTE this
estimate is formed from a sparse set of pilot (reference) symbols and then
interpolated or filtered across the time--frequency grid~\cite{coleri,vandebeek}.
The quality of this estimate directly bounds the achievable bit-error rate
(BER) and, ultimately, the link throughput, which makes channel estimation one
of the most studied receiver blocks in the wireless literature.

Decades of work have produced a spectrum of estimators trading complexity for
accuracy: LS with linear or spline interpolation~\cite{coleri}; discrete
Fourier transform (DFT) based denoising that exploits the finite delay spread
of the channel~\cite{zhaohuang}; robust LMMSE built on an assumed uniform
power-delay profile (PDP)~\cite{licimini}; rank-reduced singular-value
decomposition (SVD) LMMSE~\cite{edfors}; full frequency-domain LMMSE using the
estimated channel covariance~\cite{edfors}; joint space--frequency LMMSE for
MIMO arrays~\cite{barhumi,biguesh}; and, most recently, learned convolutional
denoisers such as ChannelNet~\cite{soltani}. A recurring---but rarely
quantified---observation is that \emph{none of these is universally best.} A
low-complexity LS estimator can rival LMMSE at moderate SNR in a
low-delay-spread channel, an LMMSE tuned to one PDP degrades under model
mismatch, DFT denoisers floor early because of spectral leakage, and learned
estimators generalize well at low SNR but saturate at high SNR. The optimal
choice migrates with the scenario, numerology, SNR, Doppler, and antenna
geometry.

This paper makes that migration explicit and then exploits it. Our
contributions are:
\begin{itemize}
\item[(i)] A \emph{unified benchmark} of eight literature estimators over the
3GPP TR~38.901 UMa/UMi/RMa channels (plus TDL-C for mobility), generated with
NVIDIA Sionna~\cite{sionna}, spanning both 5G-NR and LTE numerologies and both
SISO and $8\times2$ MIMO, and evaluated by NMSE, coded/uncoded BER, and Doppler
robustness.
\item[(ii)] A demonstration, with win-count and regret statistics, that the
best estimator changes across operating points---including a clean
low-SNR/high-SNR regime switch in the MIMO case.
\item[(iii)] A \emph{condition-adaptive multi-agent orchestrator} that
dispatches to the validation-best estimator per condition (genie-free) and
provably upper-bounds any fixed strategy, tracking the per-realization oracle to
within $1.07$~dB.
\item[(iv)] A \emph{concurrency} analysis showing that, because the agents are
independent, orchestration achieves best-of-eight accuracy at near
single-estimator latency, and that data-parallel partitioning scales as $\sim
1/K$ while by-algorithm partitioning is Amdahl-limited.
\end{itemize}

\section{System Model}
\subsection{OFDM signal model}
Consider an OFDM symbol with $N$ subcarriers. Let $\mathbf{h}\in\mathbb{C}^{N}$
be the CFR of a link and $\mathbf{H}=\mathrm{diag}(\mathbf{h})$. The received
frequency-domain symbol is
\begin{equation}
\mathbf{y} = \mathbf{H}\mathbf{x} + \mathbf{n},\qquad
\mathbf{n}\sim\mathcal{CN}(\mathbf{0},\sigma^2\mathbf{I}),
\end{equation}
where $\mathbf{x}$ carries known pilots on the comb set
$\mathcal{P}\subset\{0,\dots,N-1\}$, $|\mathcal{P}|=N_p$. Assuming unit-modulus
pilots, the LS estimate at the pilots is
$\hat{\mathbf{h}}_{\mathcal{P}}^{\mathrm{LS}}
=\mathbf{h}_{\mathcal{P}}+\tilde{\mathbf{n}}_{\mathcal{P}}$, from which the full
CFR $\hat{\mathbf{h}}\in\mathbb{C}^{N}$ is produced by each estimator. We use
$N=128$ subcarriers with comb pilots: 5G-NR uses subcarrier spacing (SCS)
$\Delta f=30$~kHz at carrier $f_c=3.5$~GHz with comb-4 pilots ($N_p=32$), and
LTE uses $\Delta f=15$~kHz at $f_c=2.1$~GHz with comb-8 pilots ($N_p=16$).

\subsection{MIMO extension}
For the MIMO study we consider a downlink $8\times2$ configuration: an
eight-element dual-polarized base-station panel ($N_t=8$) and a two-element
dual-polarized user equipment ($N_r=2$), giving $L=N_rN_t=16$ spatial links.
Stacking the per-link CFRs yields $\mathbf{h}\in\mathbb{C}^{LN}$, estimated
either per link or jointly across links (Section~\ref{sec:algos}).

\subsection{Channel models}
Channel realizations are drawn from the 3GPP TR~38.901 system-level
models---UMa, UMi and RMa~\cite{tr38901}---using the NVIDIA Sionna PyTorch
backend~\cite{sionna}. Large-scale pathloss and shadowing are disabled and each
link is power-normalized so that the SNR $=1/\sigma^2$ is well defined. For the
mobility study we additionally use the TDL-C tapped-delay model, sweeping the
maximum Doppler $f_d$ from $0$ to $1400$~Hz (equivalently $0$--$120$~km/h at
$3.5$~GHz).

\subsection{Performance metric}
The primary accuracy metric is the normalized mean-square error, pooled over
links and subcarriers,
\begin{equation}
\mathrm{NMSE}=\frac{\mathbb{E}\!\left[\|\hat{\mathbf{h}}-\mathbf{h}\|_2^2\right]}
{\mathbb{E}\!\left[\|\mathbf{h}\|_2^2\right]}.
\end{equation}
We report NMSE in dB, and for the link study we also measure the uncoded BER of
a QPSK payload against a genie (perfect-CSI) receiver.

\section{Channel Estimation Algorithms}
\label{sec:algos}
We benchmark eight estimators drawn from the literature. All share the interface
$\hat{\mathbf{h}}_{\mathcal{P}}^{\mathrm{LS}}\!\rightarrow\!\hat{\mathbf{h}}$.

\emph{LS + linear / spline}~\cite{coleri}: interpolate the pilot LS estimates
across the comb with linear or natural cubic-spline interpolation; no channel
statistics are used.

\emph{DFT-based}~\cite{zhaohuang}: transform the LS estimate to the time domain,
retain the leading $L_\tau$ taps that contain the channel energy, null the
noise-dominated tail, and transform back.

\emph{Robust-LMMSE}~\cite{licimini}: an LMMSE filter built from an assumed
\emph{uniform} PDP out to a maximum delay $\tau_{\max}$, robust to unknown
statistics. With correlation $\mathbf{R}$ the estimator is
\begin{equation}
\hat{\mathbf{h}} = \mathbf{R}_{a\mathcal{P}}\!
\left(\mathbf{R}_{\mathcal{P}\mathcal{P}}+\sigma^2\mathbf{I}\right)^{-1}
\hat{\mathbf{h}}_{\mathcal{P}}^{\mathrm{LS}}.
\label{eq:lmmse}
\end{equation}

\emph{SVD-LMMSE}~\cite{edfors}: (\ref{eq:lmmse}) restricted to the dominant
rank-$r$ eigen-subspace of the pilot covariance, reducing complexity and noise.

\emph{Freq-LMMSE}~\cite{edfors}: (\ref{eq:lmmse}) with the \emph{estimated}
per-link frequency covariance $\mathbf{R}=\mathbb{E}[\mathbf{h}\mathbf{h}^{H}]$.

\emph{SpaceFreq-LMMSE}~\cite{barhumi,biguesh}: a joint MIMO estimator that
exploits correlation \emph{across} the $L$ links as well as across frequency.
The length-$LN$ stacked channel is estimated from the $LN_p$ stacked pilot
observations using the joint covariance $\mathbf{R}\in\mathbb{C}^{LN\times LN}$
in (\ref{eq:lmmse}).

\emph{CNN}~\cite{soltani}: a residual one-dimensional convolutional denoiser
(ChannelNet-style) applied to the linearly interpolated LS estimate of each
link, trained across UMa/UMi/RMa and both numerologies with SNR randomized in
$[-5,25]$~dB.

\section{Multi-Agent Orchestration}
\label{sec:orch}
\subsection{Accuracy: condition-adaptive dispatch}
Let $\mathcal{A}=\{a_1,\dots,a_M\}$ be the set of estimator agents
($M=8$). For each operating \emph{condition} $c$ (a tuple of channel model,
numerology, SNR, and antenna configuration) we split test realizations into a
validation half and a test half. The orchestrator selects
\begin{equation}
a^\star(c)=\arg\min_{a\in\mathcal{A}}
\mathrm{NMSE}_{\mathrm{val}}(a,c),
\end{equation}
and reports $\mathrm{NMSE}_{\mathrm{test}}(a^\star(c),c)$. This uses no genie
knowledge---only the SNR and scenario label already available at the
receiver---and by construction upper-bounds any single fixed agent. As a lower
bound we also compute the per-realization \emph{oracle} that selects the best
agent for each individual channel drop; the gap between orchestrator and oracle
quantifies the residual headroom.

\subsection{Speed: independent agents run concurrently}
The agents share no state, so they can be executed in parallel. We study two
partitionings on $K$ worker processes (each pinned to a single thread):
\begin{itemize}
\item \emph{By-algorithm}: one estimator per worker. Wall-clock is
$\max_a t_a$; this is Amdahl-limited by the heaviest agent (the CNN).
\item \emph{Data-parallel}: every worker runs \emph{all} estimators on a $1/K$
shard of the channel realizations. Wall-clock scales as $\sim 1/K$.
\end{itemize}
Crucially, the orchestrator's own selection cost is the \emph{sum} of the agent
runtimes when executed sequentially by a single agent, but only the \emph{max}
when executed concurrently---so orchestration buys best-of-$M$ accuracy at
roughly single-estimator latency.

\emph{Single-realization limit.} When only one channel realization is available,
data-parallel sharding is impossible ($N=1$ cannot be split), and the only
remaining axis is the by-algorithm partition. Running the bank is then a
makespan problem over $M$ indivisible tasks whose optimum---for any number of
workers---is the heaviest single agent, $t_{\text{best}}=\max_a t_a=t_{\text{CNN}}$.
With serial fraction $s=t_{\text{CNN}}/\sum_a t_a$, Amdahl's law caps the speedup
at $1/[s+(1-s)/K]\to 1/s$ as $K\to\infty$; for our bank $s\approx0.97$, so
concurrency yields only $\approx\!1.03\times$ over sequential execution. In this
regime orchestration therefore buys \emph{accuracy}, not speed: best-of-$M$
selection at \emph{max}- rather than \emph{sum}-latency, with the heaviest agent
as a hard floor. Reducing single-realization latency further requires a
different axis---intra-agent parallelism (splitting the CNN's own computation),
a cheaper heavy agent (pruning/quantization/distillation), or a conditional
cascade that invokes the CNN only when a confidence gate deems it necessary.

\section{Tests, Simulations, and Results}
We first present the SISO link study (accuracy, BER, mobility, concurrency) and
then the $8\times2$ MIMO study (accuracy, orchestration gain, speed).

\subsection{SISO: estimation accuracy}
\begin{figure*}[!tb]
\centering
\includegraphics[width=0.85\textwidth]{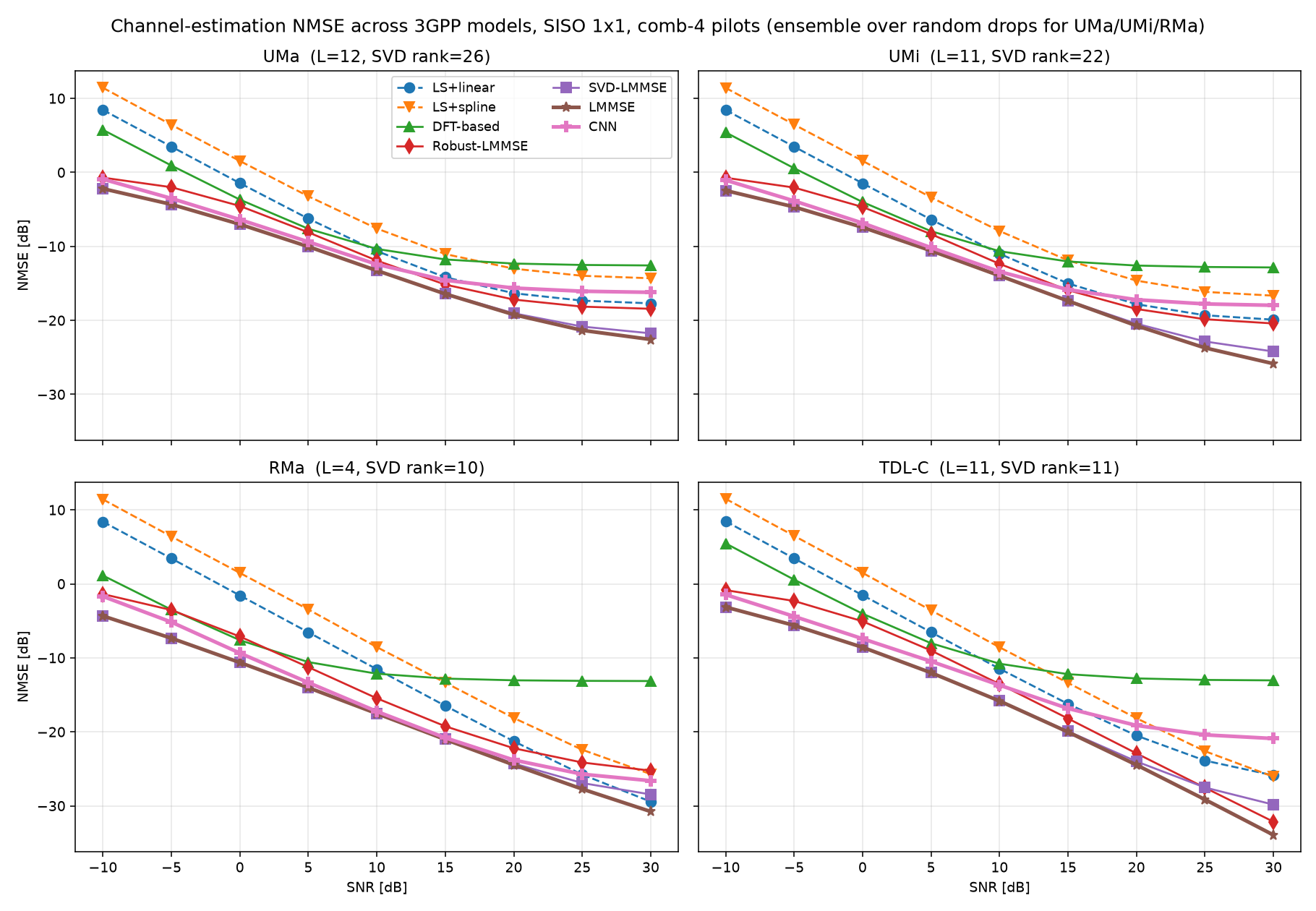}
\caption{SISO channel-estimation NMSE versus SNR for the eight estimators over
3GPP UMa/UMi/RMa and TDL-C. The best estimator and the spread between
estimators change with the channel model and SNR.}
\label{fig:siso_nmse}
\end{figure*}

Fig.~\ref{fig:siso_nmse} shows the NMSE versus SNR for all estimators over
UMa/UMi/RMa and TDL-C. The full-covariance \emph{LMMSE} and its low-rank
\emph{SVD-LMMSE} approximation are best at high SNR (e.g.\ $-22.6$ and
$-21.8$~dB at $30$~dB on UMa, and $-33.9$~dB for LMMSE on the low-spread
TDL-C), while the \emph{DFT-based} estimator floors early ($\approx-12.6$~dB) due
to spectral leakage of off-grid taps. The learned \emph{CNN} is the most
robust at low SNR ($-6.4$~dB at $0$~dB on UMa, matching LMMSE) but saturates
near $-16$~dB at high SNR because its training loss is dominated by the
noisy regime. Notably, simple \emph{LS+linear} interpolation is within
$1$--$2$~dB of LMMSE at moderate SNR, confirming that estimator ranking is not
fixed.

\subsection{SISO: bit-error rate}
\begin{figure*}[!tb]
\centering
\includegraphics[width=0.83\textwidth]{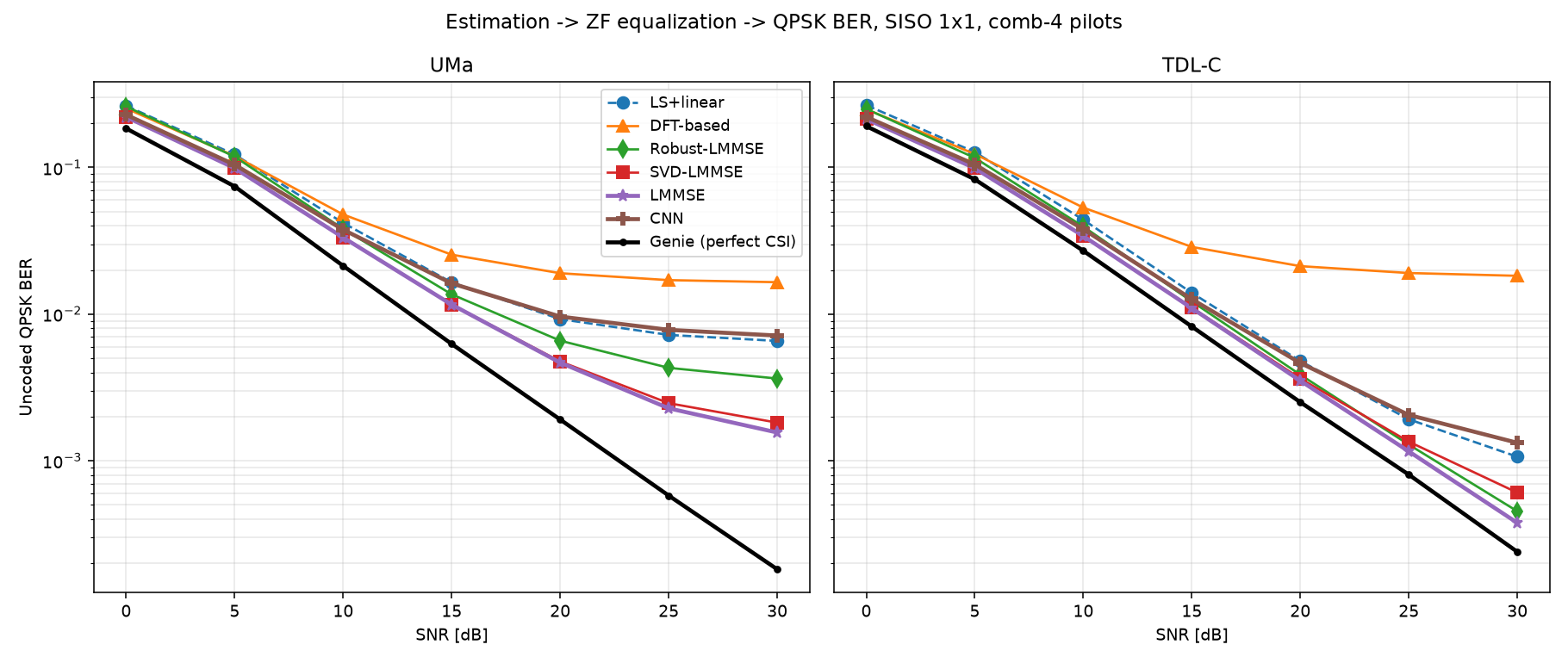}
\caption{SISO uncoded QPSK BER versus SNR (UMa) against a genie (perfect-CSI)
receiver. LMMSE/SVD-LMMSE track the genie most closely; the DFT estimator shows
an error floor.}
\label{fig:siso_ber}
\end{figure*}

Fig.~\ref{fig:siso_ber} plots uncoded QPSK BER against a genie receiver on UMa.
The accuracy ranking carries over to detection: LMMSE and SVD-LMMSE track the
genie most closely (BER $\approx1.6\times10^{-3}$ vs.\ $1.8\times10^{-4}$ at
$30$~dB), Robust-LMMSE follows, and the DFT-based estimator exhibits an error
floor ($\approx1.7\times10^{-2}$) inherited from its NMSE floor. The CNN and
LS+linear sit between, competitive at low-to-moderate SNR.

\subsection{SISO: mobility (Doppler)}
\begin{figure}[H]
\centering
\includegraphics[width=0.95\columnwidth]{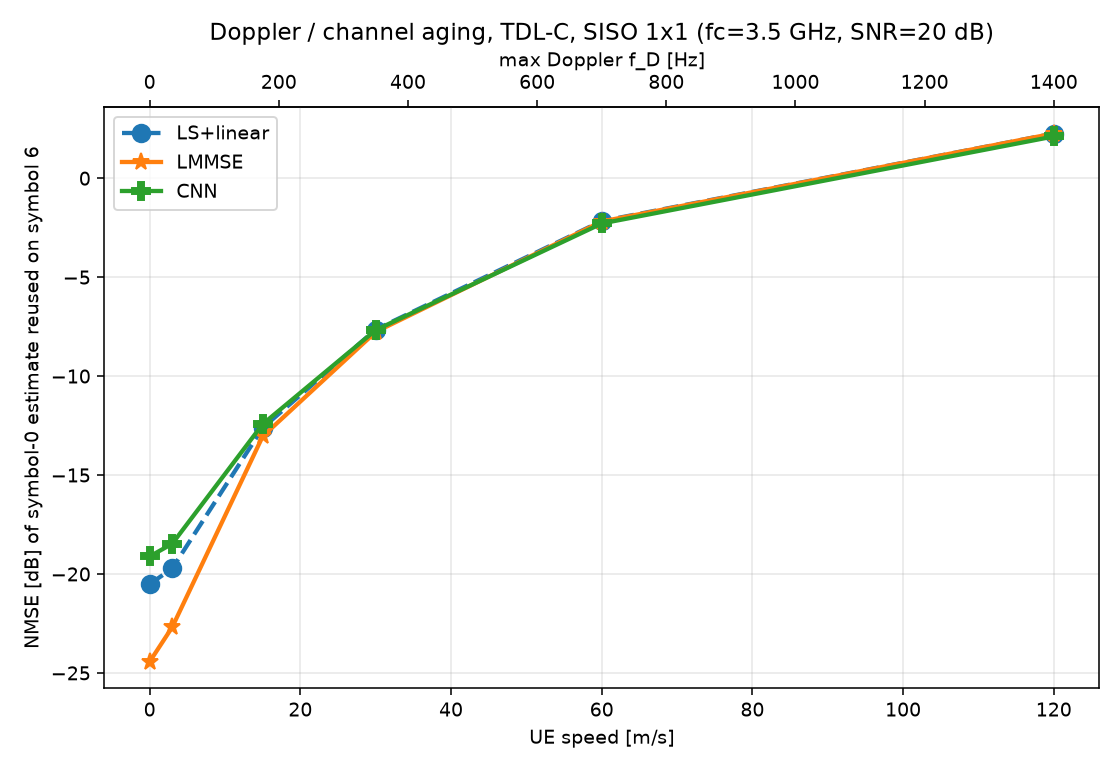}
\caption{SISO NMSE versus maximum Doppler at $20$~dB SNR (TDL-C). All estimators
degrade and converge as mobility induces intra-symbol time variation.}
\label{fig:siso_doppler}
\end{figure}

Fig.~\ref{fig:siso_doppler} sweeps the maximum Doppler at a fixed $20$~dB SNR.
At low Doppler, LMMSE leads by $\approx4$--$6$~dB; as $f_d$ grows the
inter-carrier interference from time-variation within the symbol dominates and
\emph{all} estimators converge to a poor NMSE ($\approx+2$~dB at
$f_d=1400$~Hz). This is a second axis along which the best estimator---and the
value of statistical filtering---changes.

\subsection{SISO: concurrency}
\begin{figure*}[!tb]
\centering
\includegraphics[width=0.83\textwidth]{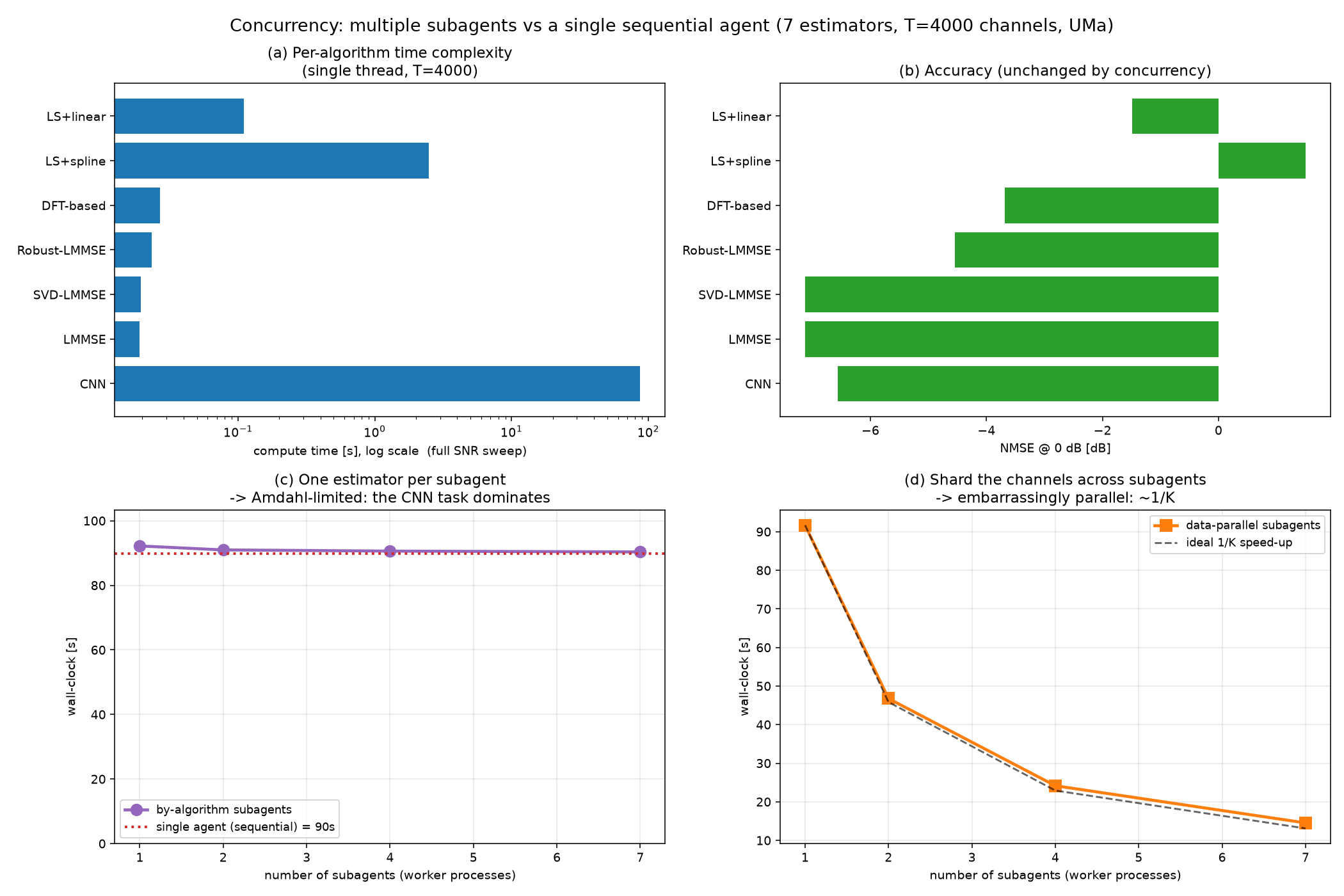}
\caption{SISO wall-clock of the estimator bank. By-algorithm partitioning is
Amdahl-limited by the CNN agent; data-parallel partitioning scales as $\sim
1/K$.}
\label{fig:siso_conc}
\end{figure*}

Fig.~\ref{fig:siso_conc} reports the wall-clock of running the estimator bank.
A single agent evaluates all seven estimators sequentially in $89.8$~s, of which
the CNN alone is $87.1$~s. By-algorithm partitioning therefore barely improves
($90.3$~s at $K=7$), a textbook Amdahl's-law ceiling. In contrast, the
data-parallel partition scales from $91.7$~s to $14.5$~s at $K=7$
($6.3\times$), close to the ideal $1/K$.

\subsection{MIMO: estimation accuracy and the regime switch}
\begin{figure*}[!tb]
\centering
\includegraphics[width=0.83\textwidth]{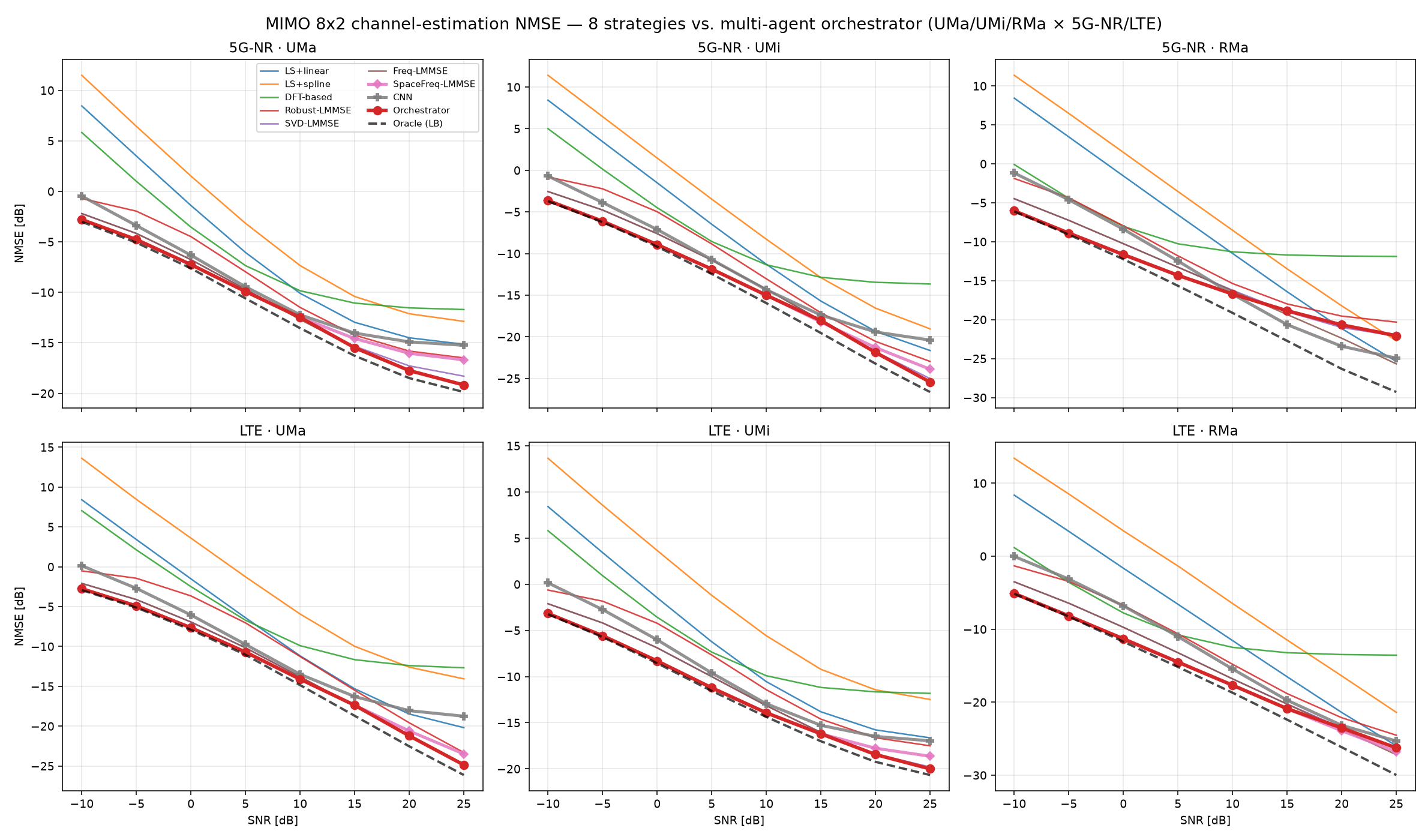}
\caption{$8\times2$ MIMO NMSE versus SNR for the eight strategies and the
multi-agent orchestrator over UMa/UMi/RMa $\times$ 5G-NR/LTE. The orchestrator
(red) hugs the per-realization oracle (dashed) and dominates every fixed
strategy.}
\label{fig:mimo_nmse}
\end{figure*}

Fig.~\ref{fig:mimo_nmse} shows the $8\times2$ MIMO NMSE over the six
condition tiles (UMa/UMi/RMa $\times$ 5G-NR/LTE). Two regimes are visible. At
\emph{low} SNR the joint \emph{SpaceFreq-LMMSE} dominates everywhere, because
correlation across the $16$ links provides an additional degree of noise
averaging. At \emph{high} SNR the per-link \emph{Freq-LMMSE} and
\emph{SVD-LMMSE} take over: once noise is small, the joint estimator's
covariance-model mismatch and rank limits cost more than the spatial gain. Over
the $48$ operating points (six conditions $\times$ eight SNRs), the ``best''
estimator is split across four different methods (Table~\ref{tab:wins}):
SpaceFreq-LMMSE wins $32$, Freq-LMMSE $11$, the CNN $3$, and SVD-LMMSE $2$.

\begin{table}[!t]
\caption{MIMO best-estimator win count over the $48$ operating points
(six conditions $\times$ eight SNRs). No single strategy is best everywhere.}
\label{tab:wins}
\centering
\begin{tabular}{lc}
\toprule
Estimator (agent) & \# operating points won \\
\midrule
SpaceFreq-LMMSE (joint MIMO) & 32 \\
Freq-LMMSE (per-link)        & 11 \\
CNN (deep learning)          & 3 \\
SVD-LMMSE (low-rank)         & 2 \\
\bottomrule
\end{tabular}
\end{table}

\subsection{MIMO: orchestration gain}
The orchestrator dispatches to the validation-best agent per condition and, as
shown in Fig.~\ref{fig:mimo_nmse}, tracks the winner across the whole grid. Its
average NMSE is $-13.49$~dB, within $1.07$~dB of the per-realization oracle
($-14.56$~dB) and better than the best \emph{fixed} strategy, SpaceFreq-LMMSE
($-13.28$~dB). The whole-grid average understates the benefit because it is
dominated by large low-SNR magnitudes; restricting to the high-SNR slice
($\geq15$~dB), where Freq-LMMSE wins every $+25$~dB point, a system frozen on the
low-SNR champion SpaceFreq-LMMSE loses up to \textbf{$3.6$~dB} (on RMa). The
orchestrator simply follows the winner into the new regime---exactly the
robustness a single fixed estimator cannot provide.

\subsection{MIMO: speed}
\begin{figure*}[!tb]
\centering
\includegraphics[width=0.83\textwidth]{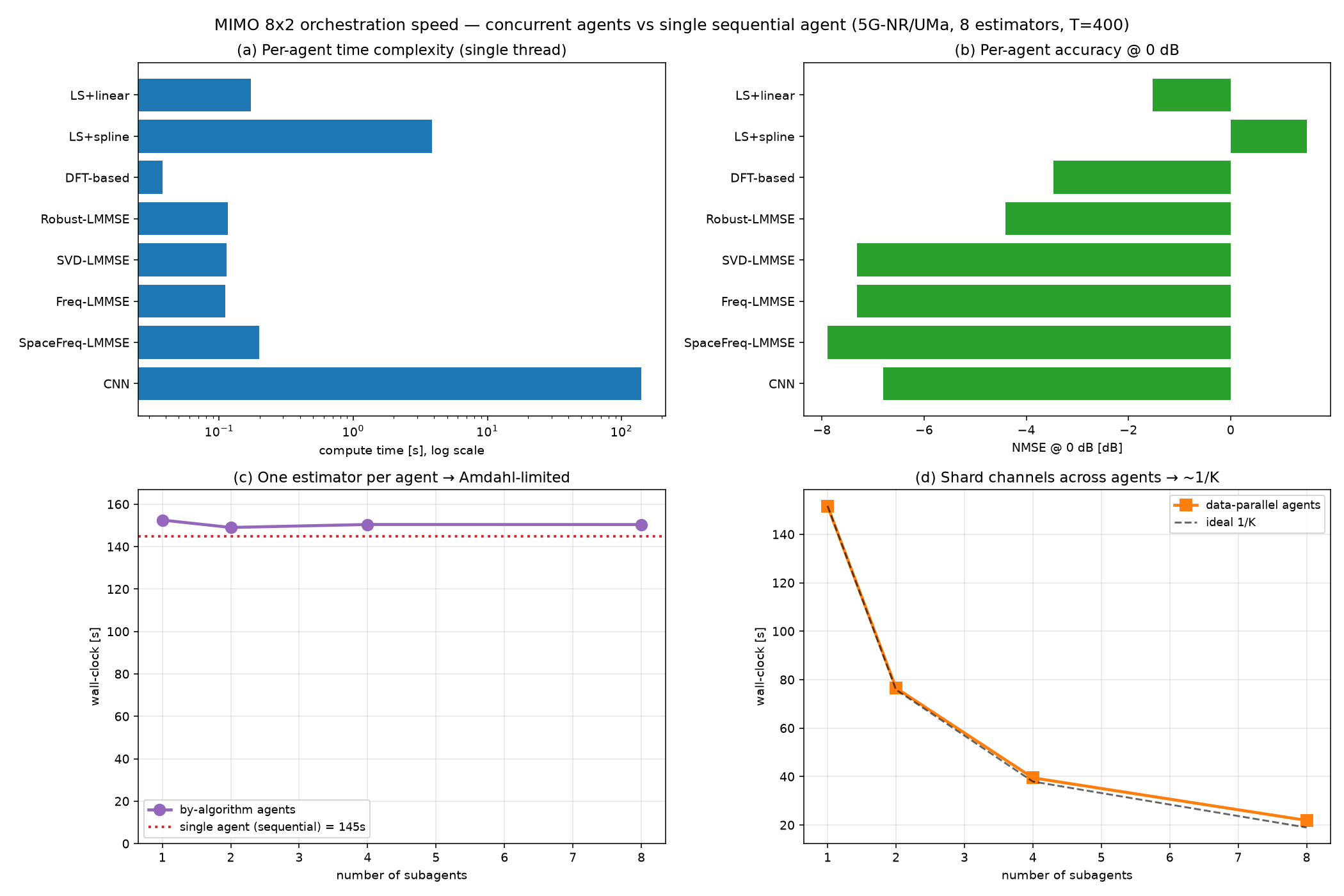}
\caption{$8\times2$ MIMO orchestration speed. (a) per-agent compute, (b)
per-agent accuracy, (c) by-algorithm partitioning is Amdahl-limited, (d)
data-parallel partitioning scales as $\sim 1/K$.}
\label{fig:mimo_orch}
\end{figure*}

Fig.~\ref{fig:mimo_orch} reports the speed benchmark on the 5G-NR/UMa condition.
A single agent runs all eight estimators sequentially in $145$~s, dominated by
the CNN ($140$~s). By-algorithm partitioning stays at $\approx150$~s for all
$K$---Amdahl-locked by the CNN agent---whereas data-parallel partitioning scales
from $152$~s to $22$~s at $K=8$ ($6.9\times$, near the ideal $1/K$). Executed
concurrently, the orchestrator pays the \emph{max} agent time ($\sim140$~s)
rather than the \emph{sum} ($\sim145$~s), delivering best-of-eight accuracy at
essentially single-estimator latency.

\subsection{Discussion}
Across both the SISO and MIMO studies, and along every axis we
varied---channel model, numerology, SNR, Doppler, and antenna
configuration---the best estimator changed. This is the empirical justification
for orchestration: rather than committing a receiver to one estimator (and its
worst-case behavior), a lightweight condition-adaptive selector achieves
near-oracle accuracy, and concurrent execution makes the selection cost
negligible. The approach is also extensible: adding a new estimator agent, or
replacing the argmin selector with a learned policy, can only improve the
orchestrator, whereas a single-strategy design is fixed at design time.

\section{Conclusion}
We presented a unified SISO-to-MIMO benchmark of eight channel estimators for
5G-NR and LTE over 3GPP TR~38.901 UMa/UMi/RMa channels generated with NVIDIA
Sionna, and showed that no single estimator is best across scenarios,
numerologies, SNR, mobility, or antenna configuration. We proposed a
condition-adaptive multi-agent orchestrator that dispatches to the
validation-best estimator per condition; it upper-bounds any fixed strategy,
tracks the per-realization oracle to within $1.07$~dB, and improves NMSE over
the best fixed strategy by up to $3.6$~dB in the high-SNR MIMO regime. Because
the estimator agents are independent, data-parallel concurrency scales the
wall-clock nearly as $1/K$ (up to $6.9\times$), so orchestration delivers
best-of-eight accuracy at near single-estimator latency. Multi-agent
orchestration is thus a practical and extensible route to robust channel
estimation across heterogeneous 5G-NR/LTE deployments.

\bibliographystyle{IEEEtran}
\bibliography{refs}

@article{coleri,
  author  = {S. Coleri and M. Ergen and A. Puri and A. Bahai},
  title   = {Channel estimation techniques based on pilot arrangement in {OFDM} systems},
  journal = {IEEE Trans. Broadcast.},
  volume  = {48},
  number  = {3},
  pages   = {223--229},
  year    = {2002},
}

@inproceedings{vandebeek,
  author    = {J.-J. van de Beek and O. Edfors and M. Sandell and S. K. Wilson and P. O. B{\"o}rjesson},
  title     = {On channel estimation in {OFDM} systems},
  booktitle = {Proc. IEEE Veh. Technol. Conf. (VTC)},
  pages     = {815--819},
  year      = {1995},
}

@inproceedings{zhaohuang,
  author    = {Y. Zhao and A. Huang},
  title     = {A novel channel estimation method for {OFDM} mobile communication systems based on pilot signals and transform-domain processing},
  booktitle = {Proc. IEEE Veh. Technol. Conf. (VTC)},
  pages     = {2089--2093},
  year      = {1997},
}

@article{licimini,
  author  = {Y. Li and L. J. Cimini and N. R. Sollenberger},
  title   = {Robust channel estimation for {OFDM} systems with rapid dispersive fading channels},
  journal = {IEEE Trans. Commun.},
  volume  = {46},
  number  = {7},
  pages   = {902--915},
  year    = {1998},
}

@article{edfors,
  author  = {O. Edfors and M. Sandell and J.-J. van de Beek and S. K. Wilson and P. O. B{\"o}rjesson},
  title   = {{OFDM} channel estimation by singular value decomposition},
  journal = {IEEE Trans. Commun.},
  volume  = {46},
  number  = {7},
  pages   = {931--939},
  year    = {1998},
}

@article{barhumi,
  author  = {I. Barhumi and G. Leus and M. Moonen},
  title   = {Optimal training design for {MIMO} {OFDM} systems in mobile wireless channels},
  journal = {IEEE Trans. Signal Process.},
  volume  = {51},
  number  = {6},
  pages   = {1615--1624},
  year    = {2003},
}

@article{biguesh,
  author  = {M. Biguesh and A. B. Gershman},
  title   = {Training-based {MIMO} channel estimation: a study of estimator tradeoffs and optimal training signals},
  journal = {IEEE Trans. Signal Process.},
  volume  = {54},
  number  = {3},
  pages   = {884--893},
  year    = {2006},
}

@article{soltani,
  author  = {M. Soltani and V. Pourahmadi and A. Mirzaei and H. Sheikhzadeh},
  title   = {Deep learning-based channel estimation},
  journal = {IEEE Commun. Lett.},
  volume  = {23},
  number  = {4},
  pages   = {652--655},
  year    = {2019},
}

@techreport{tr38901,
  author      = {{3GPP}},
  title       = {Study on channel model for frequencies from 0.5 to 100 {GHz}},
  institution = {3rd Generation Partnership Project (3GPP)},
  number      = {TR 38.901, v17.0.0},
  year        = {2022},
}

@article{sionna,
  author  = {J. Hoydis and S. Cammerer and F. {Ait Aoudia} and A. Vem and N. Binder and G. Marcus and A. Keller},
  title   = {Sionna: An open-source library for next-generation physical layer research},
  journal = {arXiv:2203.11854},
  year    = {2022},
}

\end{document}